%% file: researchnote_pra.tex
\documentclass[superscriptaddress,twocolumn, showpacs,amsmath,amssymb,aps,pra,floatfix,10pt]{revtex4-1}
\usepackage[latin1]{inputenc}
\usepackage[american]{babel}
\usepackage[T1]{fontenc}
\usepackage{mathrsfs}
\usepackage{hyperref}
\usepackage{multirow}
\usepackage{bm, epsfig}
\usepackage{graphicx}
\usepackage{subeqnarray}
\usepackage{bbold}
\usepackage{upgreek}

\include{phys_jdf}

\hypersetup{colorlinks=true, urlcolor=blue, citecolor=blue, filecolor=blue, linkcolor=blue}

\usepackage{changes}
\definechangesauthor[name={Natalia}, color=magenta]{no}
\definechangesauthor[name={Julian}, color=orange]{jcb}

\begin{document}

\title{Second-order hyperfine structure and its impact on new physics searches using isotope shift spectroscopy}

\date{20 June 2025}

\author{Julian C. Berengut}
\email[Email: ]{julian.berengut@unsw.edu.au}
\affiliation{School of Physics, University of New South Wales, Sydney NSW 2052, Australia}
\author{Natalia~S.~Oreshkina}
\email[Email: ]{natalia.oreshkina@mpi-hd.mpg.de} 
\affiliation{Max-Planck-Institut f\"ur Kernphysik, Saupfercheckweg 1, 69117 Heidelberg, Germany}

\begin{abstract}
Recent precision isotope shift spectroscopy experiments have reported King plot non-linearities that may uncover previously hidden nuclear properties or hint at new force-carrying bosons. Interpreting the data requires many isotope pairs, but sufficient stable spin-zero nuclei are not always available. In this paper we consider the use of isotopes with non-zero nuclear spin for these studies. We confirm that observed King plot non-linearity in strontium is explained by second-order hyperfine shift in the fermionic nucleus, and we discuss the limitations of using Yb isotopes with non-zero spin in King plot studies. By introducing a specific difference of two isotopes with nuclear spin, we allow for a promising addition to future experimental schemes.
\end{abstract}

\maketitle

\section{Introduction}
Precision measurements of isotope shifts have been used as a sensitive probe for hypothetical new bosons that would couple electrons and neutrons~\cite{berengut18prl,JulCed_2025,Cedric_PhysRevD.96.093001}. 
Such a boson would cause a deviation from the lowest-order mass and field shifts, which manifests as a nonlinearity in the King plot~\cite{king63josa}. 
Dedicated experiments in Ca$^+$ have seen no deviation from linearity, allowing strong limits to be placed on the coupling strength of a new boson~\cite{solaro20prl,chang_PhysRevA.110.L030801}. 
On the other hand, experiments in Yb$^+$~\cite{counts20prl,hur22prl,door25prl} and Yb~\cite{figueroa22prl,ono22prx} have seen large deviations from the linear isotope shift coming from at least two separate sources. Very recent measurements comparing the highly-charged ion Ca$^{14+}$ with Ca$^+$ have also seen large deviations from linearity~\cite{wilzewski25prl}.
It is possible to remove these sources of systematic error and extract even stronger limits on new boson couplings using the generalised King plot method~\cite{berengut20prr}, but this requires more pairs of isotopes than are currently available. 
Previous studies have almost exclusively focused on spinless nuclei because of the added complexity of the hyperfine interaction. However, in an effort to obtain additional isotopes when stable spinless isotopes are in short supply, one may consider the possibility of using nuclei with non-zero spin and averaging over the hyperfine structure.

Such an attempt has been made previously with measurements performed for strontium~\cite{miyake19prr}. 
The experiment included three spinless isotopes and one isotope with spin, for which the transition frequency was found by taking the weighted average of the three hyperfine components. 
The corresponding King plot was not a straight line, as the King theory would predict: instead a non-linearity has been observed.
Several works have speculated that this non-linearity might originate from the second-order hyperfine shift~\cite{miyake19prr,roser24pra}, as has been observed in different atomic transitions, e.g.~\cite{schelfhout21pra,Zhang_PhysRevResearch.6.043106}, but until now this hypothesis has not been confirmed.

In this paper we have calculated the second-order hyperfine shift in ${}^{87}$Sr to resolve this non-linearity. We then consider the possibility of including isotopes with spin into existing studies, and analyze the limitations of such an inclusion caused by theory accuracy. Finally, we suggest a method to overcome this difficulty by considering a specific difference of two isotopes with spin.

\section{Strontium King plot}
The frequency shift of a line $\alpha$ between isotopes with mass number $A$ and $A'$ can be expressed to good accuracy as
\begin{equation}
\label{eq:isotope_shift}
\delta\nu_\alpha^{A',A}  = F_{\alpha} \delta\langle r^2 \rangle^{A',A} +  K_\alpha \mu^{A',A}.
\end{equation}
Here $\mu^{A',A} = 1/m^{A'} - 1/m^{A}$ is the inverse nuclear mass difference, and $F$ and $K$ are the isotope-independent electronic coefficients of the field and mass shift, respectively.
By canceling the relatively poorly known difference in the mean-square nuclear charge radius between the isotopes, $\delta\langle r^2 \rangle^{A',A}$, the King plot gives a linear relationship between the modified isotope shifts ${\delta\nu_\beta^{A',A}}/{\mu^{A',A}}$ of two different transitions $\alpha$ and $\beta$:
\begin{equation}
\label{eq:kp}
\frac{\delta\nu_\beta^{A',A}}{\mu^{A',A}}  =  \frac{F_\beta}{F_\alpha} \frac{\delta\nu_\alpha^{A',A}}{\mu^{A',A}}  + \left(K_\beta-\frac{F_\beta}{F_\alpha}K_\alpha\right).
\end{equation}

The experimentally obtained King plot was presented for strontium, using isotope shifts from $A' = 84$, 86, and 87 to the reference isotope $A = 88$~\cite{miyake19prr}, see Fig. 1. 
In the experiment, the following two different transitions have been measured:  
$$\alpha \,\, (^1S_0 \rightarrow\ ^3P_1) \text{ at 689\,nm,}$$
and 
$$\beta \,\, (^1S_0 \rightarrow\ ^3P_0) \text{ at 698\,nm}.$$ 
The $^{87}$Sr nucleus has spin $I = 9/2$, and so the isotope shift $\delta\nu_{\alpha}^{87,88}$ was found by taking the weighted average of the three hyperfine components. According to the King theory \eqref{eq:isotope_shift} the three points should lie on a straight line, but they do not. 

\begin{figure}[tb]
    \centering
\includegraphics[width=\columnwidth]{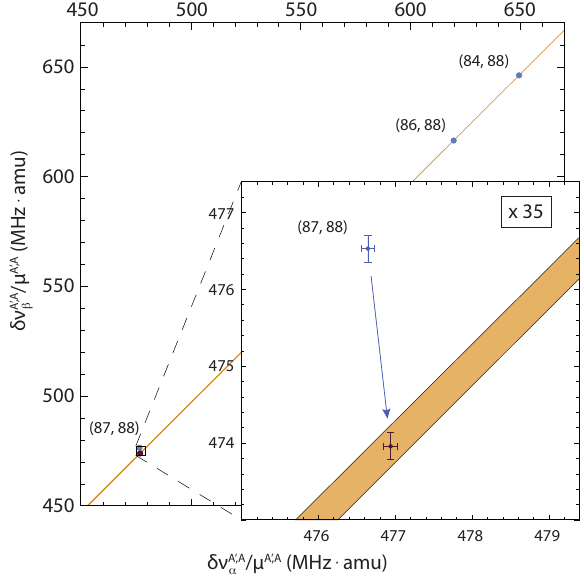}
    \caption{Strontium King plot. Blue crosses: experimental data with $1\sigma$ errors taken from~\cite{miyake19prr}, showing a large non-linearity. The orange line of best fit is based on the two upper points formed from the three bosonic isotopes. Purple cross: the experimental $^{87}$Sr 
 -- $^{88}$Sr isotope shifts, $\delta\nu^{87,88}$, are corrected by removing the calculated second-order hyperfine structure, restoring King plot linearity.}
    \label{fig:Sr_nohfs}
\end{figure}

\section{Second-order hyperfine shift}
To resolve this non-linearity, we have calculated the second-order hyperfine shift (HFS) in $^{87}$Sr. This shift is dominated by interactions within the fine-structure $3P_J$ multiplet, and in this work we only include these levels.
Formulas for the second-order HFS can be found in, e.g.~\cite{beloy08pra1, beloy08pra}, and read
\begin{multline}\label{eq:2hfs}
    {\Delta E_F^{(2)}} =   \sum_{J'}\biggl(   \left\{%
\begin{array}{c c c}
 F & J & I\\ 1 & I & J' \\
\end{array}%
\right\}^2 \mathcal{M}^2 \eta_{M1M1}  \\   + \left\{%
\begin{array}{c c c}
 F & J & I\\ 1 & I & J' \\
\end{array}%
\right\}\left\{%
\begin{array}{c c c}
 F & J & I\\ 2 & I & J' \\
\end{array}%
\right\} \mathcal{M}Q\, \eta_{M1E2} \\
+ \left\{%
\begin{array}{c c c}
 F & J & I\\ 2 & I & J' \\
\end{array}%
\right\}^2 Q^2\, \eta_{E2E2}\biggr).
\end{multline}
Here $\mathcal{M}$~\footnote{We use the notation $\mathcal{M}$ instead of the usual standard $\mu$ for the nuclear magnetic moment to avoid confusion with the reduced mass introduced in Eq.~\eqref{eq:kp}.} and $Q$ are magnetic dipole and electric quadrupole moments of the nucleus, respectively, and the $\eta$ are electronic factors:
\begin{align}
\eta_{M1M1} &= \frac{(I+1)(2I+1)}{I}
    \frac{|\langle \gamma J || T_1^e || \gamma J' \rangle|^2 }{E_{\gamma J} - E_{\gamma J'}}, \label{eq:etaMM} \\
\eta_{M1E2} &= \frac{(I+1)(2I+1)}{I}\sqrt{\frac{2I+3}{2I-1}} \notag \\
 &\quad \times \frac{\langle \gamma J || T_1^e || \gamma J' \rangle \langle \gamma J' || T_2^e || \gamma J \rangle}{E_{\gamma J} - E_{\gamma J'}}, \notag \\
\eta_{E2E2} &= \frac{(I+1)(2I+1)(2I+3)}{4I(2I-1)}
    \frac{|\langle \gamma J || T_2^e || \gamma J' \rangle|^2 }{E_{\gamma J} - E_{\gamma J'}}. \notag
\end{align}

We have calculated the off-diagonal hyperfine matrix elements such as appear in \eqref{eq:etaMM} using AMBiT~\cite{kahl19cpc}, following the CI+MBPT methods of calculating hyperfine structure presented in \cite{beloy08pra, porsev99pra}. Our basis was formed by diagonalising a set of B-splines over the Dirac-Fock core formed in the $V^{N-2}$ potential. Core correlations were  included by creating a correlation potential, which we then added to the $V^{N-2}$ potential to form Brueckner orbitals~\cite{dzuba87jpb}. Valence-valence correlations were included using the configuration interaction procedure. Matrix elements of the hyperfine structure were calculated from the many-body wavefunctions, including corrections due to the random-phase approximation. Our hyperfine $A$ and $B$ constants were found to be within 1\% and 3\% of experimental values, respectively, which gives us an estimate of the accuracy of the off-diagonal matrix elements. We included both the magnetic dipole and electric quadrupole interaction, although as expected, the former was dominant.

\section{Results and discussion}
In Table \ref{tab:results} we show the individual contributions to the second order HFS in the $^3P_J$ multiplet of Sr, while in Fig. 1 we reproduce the original King plot of~\cite{miyake19prr} and show the effect of removing the calculated second-order hyperfine shifts from the isotope shifts of the fermionic $^{87}$Sr. 
The weighted shifts are 34~kHz for the $\alpha$ transition and $-322$~kHz for $\beta$. 
Note the large shift for the $^3P_0$ level despite this level having no first-order shift. Removing these second-order hyperfine shifts completely restores the King plot linearity, enabling the extraction of tighter bounds on new physics.

\begin{table}
\begin{ruledtabular}
\begin{tabular}{cccccc}
 $J$ & $F$ & \multicolumn{4}{c}{$\Delta E_F^{(2)}$ (kHz)} \\
 & & M1M1 & M1E2 & E2E2 & Total \\
\hline
$0$ & $9/2$ & -322 & 0 & -0.062 & -322 \\
$1$ & $7/2$ & -51.3 & 4.88 & -0.116 & -47 \\
    & $9/2$ & 237 & 2.03 & -0.012 & 239 \\
    & $11/2$ & -77.9 & -4.94 & -0.078 & -83 \\
$2$ & $5/2$ & 0 & 0 & 0 & 0 \\
    & $7/2$ & 51.3 & -4.88 & 0.116 & 47 \\
    & $9/2$ & 85.2 & -2.03 & 0.074 & 83 \\
    & $11/2$ & 77.9 & 4.94 & 0.078 & 83\\
    & $13/2$ & 0 & 0 & 0 & 0 \\
\end{tabular}
\end{ruledtabular}
\caption{Contributions to the second-order hyperfine shifts of levels in the Sr $^3P$ manifold.}
\label{tab:results}
\end{table}

Note that similar calculations were also recently performed for the $5s5p\ ^3P_0 \rightarrow 5s6d\ ^3D_1$ transition in Sr~\cite{Zhang_PhysRevResearch.6.043106}, who also achieved King-plot linearity by removing second-order HFS within the $^3D$ multiplet. 
The effect in the $^3P$ multiplet, which is responsible for the nonlinearity in~\cite{miyake19prr}, seems to not have been considered in~\cite{Zhang_PhysRevResearch.6.043106} since it is two orders-of-mangitude smaller than the $^3D_1$ shift and is therefore well within the uncertainty of their calculation.

Our success in restoring King plot linearity in Sr encourages us to consider the possibility of including a well-studied, stable fermionic Yb isotope such as $^{171}$Yb~\cite{godun14prl,huntemann16prl} into the existing studies. Calculations are considerably more difficult for Yb and Yb$^+$, however we have performed calculations using the method described above for those transitions that have been measured to within a few Hz~\cite{ono22prx,door25prl}. In Yb$^+$ the fine-structure spacings are relatively large, making the corresponding second-order hyperfine shifts somewhat smaller than in Sr: $\sim$10 kHz for the $5d_{3/2}\ (F=2)$ level, and around 0.2~kHz for the $4f_{7/2}^{-1}\ 6s^2\ (F=3)$ level. There is also considerable configuration mixing in Yb$^+$, so more precise calculations should also consider the contribution of levels outside of the fine-structure manifold. On the other hand, we calculate a relatively large $-520$~kHz shift for the $^1S_0 \rightarrow\ ^3P_0$ transition in Yb, which is consistent with the value $-537\,(44)$~kHz calculated in~\cite{schelfhout21pra}. Even if this shift is calculated to within 1\% --- an ambitious goal --- the corresponding theoretical uncertainty would be three orders of magnitude larger than the current experimental errors~\cite{ono22prx}. At this point we might conclude that it is unlikely that inclusion of fermionic isotopes would improve the situation in Yb, and it seems like future developments in searches for new bosons using precision isotope shift spectroscopy may only be possible using radioactive spin-zero nuclei (e.g.~\cite{yang23ppnp}).

\section{Specific difference for isotopes with spin}
In some cases, however, we may be able to remove the second-order hyperfine structure either partially or completely in a data-driven way without requiring extremely accurate calculations. In the second-order HFS contributions shown in  Eq.~\eqref{eq:2hfs} the $\eta$ factors, along with the field and mass shift constants in Eq.~\eqref{eq:kp}, are almost isotope independent. 
For simplicity, let us consider the case when only first term in Eq.~\eqref{eq:2hfs} contributes,
\begin{align}
    \Delta E^{(2)}_{F}=\left\{%
\begin{array}{c c c}
 F & J & I\\ 1 & I & J' \\
\end{array}%
\right\}^2 \mathcal{M}^2\,\eta_{M1M1},
\end{align}
which can be calculated, as we mentioned before, with 1\% accuracy in the best case. If the theoretical accuracy is comparable to or better than the experimental accuracy, we can simply subtract this term from the corresponding transition energy and go with the King plot analysis as usual.
If the theoretical uncertainties are larger than the experimental one, we can consider a specific difference of hyperfine-averaged transition frequencies of two isotopes with spin, $A'$ and $A''$, with respect to a reference isotope $A$:
\begin{multline}
\label{eq:specific_diff}
\delta\nu_\alpha^{A',A} - \left(\frac{\mathcal{M}^{A'}}{\mathcal{M}^{A''}}\right)^2\delta\nu_\alpha^{A'',A} \\
    = F_\alpha \left( \delta\langle r^2 \rangle^{A',A}- \left(\frac{\mathcal{M}^{A'}}{\mathcal{M}^{A''}}\right)^2 \delta\langle r^2 \rangle^{A'',A} \right) \\
    + K_\alpha \left( \mu^{A',A}- \left(\frac{\mathcal{M}^{A'}}{\mathcal{M}^{A''}}\right)^2 \mu^{A'',A} \right)
\end{multline}
In this specific difference the second-order HFS cancels out, making the structure the same as for other isotope shifts. Equation~\eqref{eq:specific_diff} only relies on the ratio of nuclear magnetic moments, $\mathcal{M}^{A'}/\mathcal{M}^{A''}$, which generally can be extracted with much higher accuracy than the values themselves. The specific difference can therefore can be included within existing paradigms of searches for new physics, for example in a three-dimensional King plot or generalised King plot~\cite{berengut20prr}.
If both magnetic moments are identical, the specific difference reduces to the usual King plot difference, reflecting the cancellation of the second-order HFS in that case.

The situation is more complicated if we have two or more terms, either because of the presence of a quadrupole shift or because of the additional terms in nucleus-electron coupling and the corresponding 6j symbols. 
In such cases, one may still aim to cancel the dominant contribution corresponding to the magnetic-magnetic interaction, and the quality of the second-order HFS cancellation will depend on the ratio of dominant/subdominant contribution. 
Therefore, every case should be considered independently, and generally experiments should favor isotopes with no (or possibly small) quadrupole hyperfine shifts.

As a particular example, we consider cadmium since it is the subject of current experimental efforts~\cite{ohayon22njp}.
Cadmium has two stable isotopes with $I=1/2$, $^{111}$Cd and $^{113}$Cd, with nuclear magnetic moments of $-0.5940(3)$ and $-0.6213(3)$, respectively~\cite{IAEA_Nuclear_Moments}. Spin half isotopes do not permit a quadrupole moment, therefore the second-order HFS is defined by the M1M1 interaction and the level structure will be common for the two isotopes.
It is worth noting that while measuring two isotopes with spin will give only one point on King plot, every additional fermionic isotope will provide an another point.

\section{Conclusions}
We have explored a possibility of using isotopes with spin for the search of the new physics using King plot approach, in general cases and on specific examples. 
We showed that observed King plot non-linearity in strontium can be fully explained by second-order hyperfine
shift in the fermionic nucleus. 
Further, we studied a specific case of Yb to show that the accuracy of theoretical predictions for the second-order HFS shifts is much larger than experimental accuracy and therefore adding a isotope with a spin to King plot would not provide any additional information. Finally, we showed how a specific difference of two isotopes with spin half can be used to create an additional transition that is free of second-order HFS.
The method may be extended to other nuclei with spin, but the analysis will generally be more complex and limitations from higher-order terms should be carefully considered in every individual case.

\acknowledgements
N. S. O. thanks the Gordon Godfrey fund for financial support for her visits to UNSW Sydney, Australia.

\bibliography{ref}

\end{document}

%% file: phys_jdf.tex
\def\cpc{Comput. Phys. Commun.}

\def\josa{J. Opt. Soc. Am.}

\def\jpb{J. Phys. B}

\def\njp{New J. Phys.}

\def\pra{Phys. Rev. A}

\def\prl{Phys. Rev. Lett.}
\def\prr{Phys. Rev. Res.}
\def\prx{Phys. Rev. X}

%% file: researchnote_pra.bbl
\begin{thebibliography}{28}%
\makeatletter
\providecommand \@ifxundefined [1]{%
 \@ifx{#1\undefined}
}%
\providecommand \@ifnum [1]{%
 \ifnum #1\expandafter \@firstoftwo
 \else \expandafter \@secondoftwo
 \fi
}%
\providecommand \@ifx [1]{%
 \ifx #1\expandafter \@firstoftwo
 \else \expandafter \@secondoftwo
 \fi
}%
\providecommand \natexlab [1]{#1}%
\providecommand \enquote  [1]{``#1''}%
\providecommand \bibnamefont  [1]{#1}%
\providecommand \bibfnamefont [1]{#1}%
\providecommand \citenamefont [1]{#1}%
\providecommand \href@noop [0]{\@secondoftwo}%
\providecommand \href [0]{\begingroup \@sanitize@url \@href}%
\providecommand \@href[1]{\@@startlink{#1}\@@href}%
\providecommand \@@href[1]{\endgroup#1\@@endlink}%
\providecommand \@sanitize@url [0]{\catcode `\\12\catcode `\$12\catcode
  `\&12\catcode `\#12\catcode `\^12\catcode `\_12\catcode `\%12\relax}%
\providecommand \@@startlink[1]{}%
\providecommand \@@endlink[0]{}%
\providecommand \url  [0]{\begingroup\@sanitize@url \@url }%
\providecommand \@url [1]{\endgroup\@href {#1}{\urlprefix }}%
\providecommand \urlprefix  [0]{URL }%
\providecommand \Eprint [0]{\href }%
\providecommand \doibase [0]{http://dx.doi.org/}%
\providecommand \selectlanguage [0]{\@gobble}%
\providecommand \bibinfo  [0]{\@secondoftwo}%
\providecommand \bibfield  [0]{\@secondoftwo}%
\providecommand \translation [1]{[#1]}%
\providecommand \BibitemOpen [0]{}%
\providecommand \bibitemStop [0]{}%
\providecommand \bibitemNoStop [0]{.\EOS\space}%
\providecommand \EOS [0]{\spacefactor3000\relax}%
\providecommand \BibitemShut  [1]{\csname bibitem#1\endcsname}%
\let\auto@bib@innerbib\@empty
\bibitem [{\citenamefont {Berengut}\ \emph {et~al.}(2018)\citenamefont
  {Berengut}, \citenamefont {Budker}, \citenamefont {Delaunay}, \citenamefont
  {Flambaum}, \citenamefont {Frugiuele}, \citenamefont {Fuchs}, \citenamefont
  {Grojean}, \citenamefont {Harnik}, \citenamefont {Ozeri}, \citenamefont
  {Perez},\ and\ \citenamefont {Soreq}}]{berengut18prl}%
  \BibitemOpen
  \bibfield  {author} {\bibinfo {author} {\bibfnamefont {J.~C.}\ \bibnamefont
  {Berengut}}, \bibinfo {author} {\bibfnamefont {D.}~\bibnamefont {Budker}},
  \bibinfo {author} {\bibfnamefont {C.}~\bibnamefont {Delaunay}}, \bibinfo
  {author} {\bibfnamefont {V.~V.}\ \bibnamefont {Flambaum}}, \bibinfo {author}
  {\bibfnamefont {C.}~\bibnamefont {Frugiuele}}, \bibinfo {author}
  {\bibfnamefont {E.}~\bibnamefont {Fuchs}}, \bibinfo {author} {\bibfnamefont
  {C.}~\bibnamefont {Grojean}}, \bibinfo {author} {\bibfnamefont
  {R.}~\bibnamefont {Harnik}}, \bibinfo {author} {\bibfnamefont
  {R.}~\bibnamefont {Ozeri}}, \bibinfo {author} {\bibfnamefont
  {G.}~\bibnamefont {Perez}}, \ and\ \bibinfo {author} {\bibfnamefont
  {Y.}~\bibnamefont {Soreq}},\ }\href@noop {} {\bibfield  {journal} {\bibinfo
  {journal} {\prl}\ }\textbf {\bibinfo {volume} {120}},\ \bibinfo {pages}
  {091801} (\bibinfo {year} {2018})}\BibitemShut {NoStop}%
\bibitem [{\citenamefont {Berengut}\ and\ \citenamefont
  {Delaunay}(2025)}]{JulCed_2025}%
  \BibitemOpen
  \bibfield  {author} {\bibinfo {author} {\bibfnamefont {J.~C.}\ \bibnamefont
  {Berengut}}\ and\ \bibinfo {author} {\bibfnamefont {C.}~\bibnamefont
  {Delaunay}},\ }\href {\doibase 10.1038/s42254-024-00793-2} {\bibfield
  {journal} {\bibinfo  {journal} {Nature Reviews Physics}\ }\textbf {\bibinfo
  {volume} {7}},\ \bibinfo {pages} {119} (\bibinfo {year} {2025})}\BibitemShut
  {NoStop}%
\bibitem [{\citenamefont {Delaunay}\ \emph {et~al.}(2017)\citenamefont
  {Delaunay}, \citenamefont {Ozeri}, \citenamefont {Perez},\ and\ \citenamefont
  {Soreq}}]{Cedric_PhysRevD.96.093001}%
  \BibitemOpen
  \bibfield  {author} {\bibinfo {author} {\bibfnamefont {C.}~\bibnamefont
  {Delaunay}}, \bibinfo {author} {\bibfnamefont {R.}~\bibnamefont {Ozeri}},
  \bibinfo {author} {\bibfnamefont {G.}~\bibnamefont {Perez}}, \ and\ \bibinfo
  {author} {\bibfnamefont {Y.}~\bibnamefont {Soreq}},\ }\href {\doibase
  10.1103/PhysRevD.96.093001} {\bibfield  {journal} {\bibinfo  {journal} {Phys.
  Rev. D}\ }\textbf {\bibinfo {volume} {96}},\ \bibinfo {pages} {093001}
  (\bibinfo {year} {2017})}\BibitemShut {NoStop}%
\bibitem [{\citenamefont {King}(1963)}]{king63josa}%
  \BibitemOpen
  \bibfield  {author} {\bibinfo {author} {\bibfnamefont {W.~H.}\ \bibnamefont
  {King}},\ }\href@noop {} {\bibfield  {journal} {\bibinfo  {journal} {\josa}\
  }\textbf {\bibinfo {volume} {53}},\ \bibinfo {pages} {638} (\bibinfo {year}
  {1963})}\BibitemShut {NoStop}%
\bibitem [{\citenamefont {Solaro}\ \emph {et~al.}(2020)\citenamefont {Solaro},
  \citenamefont {Meyer}, \citenamefont {Fisher}, \citenamefont {Berengut},
  \citenamefont {Fuchs},\ and\ \citenamefont {Drewsen}}]{solaro20prl}%
  \BibitemOpen
  \bibfield  {author} {\bibinfo {author} {\bibfnamefont {C.}~\bibnamefont
  {Solaro}}, \bibinfo {author} {\bibfnamefont {S.}~\bibnamefont {Meyer}},
  \bibinfo {author} {\bibfnamefont {K.}~\bibnamefont {Fisher}}, \bibinfo
  {author} {\bibfnamefont {J.~C.}\ \bibnamefont {Berengut}}, \bibinfo {author}
  {\bibfnamefont {E.}~\bibnamefont {Fuchs}}, \ and\ \bibinfo {author}
  {\bibfnamefont {M.}~\bibnamefont {Drewsen}},\ }\href@noop {} {\bibfield
  {journal} {\bibinfo  {journal} {\prl}\ }\textbf {\bibinfo {volume} {125}},\
  \bibinfo {pages} {123003} (\bibinfo {year} {2020})}\BibitemShut {NoStop}%
\bibitem [{\citenamefont {Chang}\ \emph {et~al.}(2024)\citenamefont {Chang},
  \citenamefont {Awazi}, \citenamefont {Berengut}, \citenamefont {Fuchs},\ and\
  \citenamefont {Doret}}]{chang_PhysRevA.110.L030801}%
  \BibitemOpen
  \bibfield  {author} {\bibinfo {author} {\bibfnamefont {T.~T.}\ \bibnamefont
  {Chang}}, \bibinfo {author} {\bibfnamefont {B.~B.}\ \bibnamefont {Awazi}},
  \bibinfo {author} {\bibfnamefont {J.~C.}\ \bibnamefont {Berengut}}, \bibinfo
  {author} {\bibfnamefont {E.}~\bibnamefont {Fuchs}}, \ and\ \bibinfo {author}
  {\bibfnamefont {S.~C.}\ \bibnamefont {Doret}},\ }\href {\doibase
  10.1103/PhysRevA.110.L030801} {\bibfield  {journal} {\bibinfo  {journal}
  {Phys. Rev. A}\ }\textbf {\bibinfo {volume} {110}},\ \bibinfo {pages}
  {L030801} (\bibinfo {year} {2024})}\BibitemShut {NoStop}%
\bibitem [{\citenamefont {Counts}\ \emph {et~al.}(2020)\citenamefont {Counts},
  \citenamefont {Hur}, \citenamefont {\protect{Aude Craik}}, \citenamefont
  {Jeon}, \citenamefont {Leung}, \citenamefont {Berengut}, \citenamefont
  {Geddes}, \citenamefont {Kawasaki}, \citenamefont {Jhe},\ and\ \citenamefont
  {Vuleti\'c}}]{counts20prl}%
  \BibitemOpen
  \bibfield  {author} {\bibinfo {author} {\bibfnamefont {I.}~\bibnamefont
  {Counts}}, \bibinfo {author} {\bibfnamefont {J.}~\bibnamefont {Hur}},
  \bibinfo {author} {\bibfnamefont {D.~P.~L.}\ \bibnamefont {\protect{Aude
  Craik}}}, \bibinfo {author} {\bibfnamefont {H.}~\bibnamefont {Jeon}},
  \bibinfo {author} {\bibfnamefont {C.}~\bibnamefont {Leung}}, \bibinfo
  {author} {\bibfnamefont {J.~C.}\ \bibnamefont {Berengut}}, \bibinfo {author}
  {\bibfnamefont {A.}~\bibnamefont {Geddes}}, \bibinfo {author} {\bibfnamefont
  {A.}~\bibnamefont {Kawasaki}}, \bibinfo {author} {\bibfnamefont
  {W.}~\bibnamefont {Jhe}}, \ and\ \bibinfo {author} {\bibfnamefont
  {V.}~\bibnamefont {Vuleti\'c}},\ }\href@noop {} {\bibfield  {journal}
  {\bibinfo  {journal} {\prl}\ }\textbf {\bibinfo {volume} {125}},\ \bibinfo
  {pages} {123002} (\bibinfo {year} {2020})}\BibitemShut {NoStop}%
\bibitem [{\citenamefont {Hur}\ \emph {et~al.}(2022)\citenamefont {Hur},
  \citenamefont {\protect{Aude Craik}}, \citenamefont {Counts}, \citenamefont
  {Knyazev}, \citenamefont {Caldwell}, \citenamefont {Leung}, \citenamefont
  {Pandey}, \citenamefont {Berengut}, \citenamefont {Geddes}, \citenamefont
  {Nazarewicz}, \citenamefont {Reinhard}, \citenamefont {Kawasaki},
  \citenamefont {Jeon}, \citenamefont {Jhe},\ and\ \citenamefont
  {Vuleti\'c}}]{hur22prl}%
  \BibitemOpen
  \bibfield  {author} {\bibinfo {author} {\bibfnamefont {J.}~\bibnamefont
  {Hur}}, \bibinfo {author} {\bibfnamefont {D.~P.~L.}\ \bibnamefont
  {\protect{Aude Craik}}}, \bibinfo {author} {\bibfnamefont {I.}~\bibnamefont
  {Counts}}, \bibinfo {author} {\bibfnamefont {E.}~\bibnamefont {Knyazev}},
  \bibinfo {author} {\bibfnamefont {L.}~\bibnamefont {Caldwell}}, \bibinfo
  {author} {\bibfnamefont {C.}~\bibnamefont {Leung}}, \bibinfo {author}
  {\bibfnamefont {S.}~\bibnamefont {Pandey}}, \bibinfo {author} {\bibfnamefont
  {J.~C.}\ \bibnamefont {Berengut}}, \bibinfo {author} {\bibfnamefont
  {A.}~\bibnamefont {Geddes}}, \bibinfo {author} {\bibfnamefont
  {W.}~\bibnamefont {Nazarewicz}}, \bibinfo {author} {\bibfnamefont {P.-G.}\
  \bibnamefont {Reinhard}}, \bibinfo {author} {\bibfnamefont {A.}~\bibnamefont
  {Kawasaki}}, \bibinfo {author} {\bibfnamefont {H.}~\bibnamefont {Jeon}},
  \bibinfo {author} {\bibfnamefont {W.}~\bibnamefont {Jhe}}, \ and\ \bibinfo
  {author} {\bibfnamefont {V.}~\bibnamefont {Vuleti\'c}},\ }\href@noop {}
  {\bibfield  {journal} {\bibinfo  {journal} {\prl}\ }\textbf {\bibinfo
  {volume} {128}},\ \bibinfo {pages} {163201} (\bibinfo {year}
  {2022})}\BibitemShut {NoStop}%
\bibitem [{\citenamefont {Door}\ \emph {et~al.}(2025)\citenamefont {Door},
  \citenamefont {Yeh}, \citenamefont {Heinz}, \citenamefont {Kirk},
  \citenamefont {Lyu}, \citenamefont {Miyagi}, \citenamefont {Berengut},
  \citenamefont {Biero{\'n}}, \citenamefont {Blaum}, \citenamefont {Dreissen},
  \citenamefont {Eliseev}, \citenamefont {Filianin}, \citenamefont {Filzinger},
  \citenamefont {Fuchs}, \citenamefont {F{\"u}rst}, \citenamefont {Gaigalas},
  \citenamefont {Harman}, \citenamefont {Herkenhoff}, \citenamefont
  {Huntemann}, \citenamefont {Keitel}, \citenamefont {Kromer}, \citenamefont
  {Lange}, \citenamefont {Rischka}, \citenamefont {Schweiger}, \citenamefont
  {Schwenk}, \citenamefont {Shimizu},\ and\ \citenamefont
  {Mehlst{\"a}ubler}}]{door25prl}%
  \BibitemOpen
  \bibfield  {author} {\bibinfo {author} {\bibfnamefont {M.}~\bibnamefont
  {Door}}, \bibinfo {author} {\bibfnamefont {C.-H.}\ \bibnamefont {Yeh}},
  \bibinfo {author} {\bibfnamefont {M.}~\bibnamefont {Heinz}}, \bibinfo
  {author} {\bibfnamefont {F.}~\bibnamefont {Kirk}}, \bibinfo {author}
  {\bibfnamefont {C.}~\bibnamefont {Lyu}}, \bibinfo {author} {\bibfnamefont
  {T.}~\bibnamefont {Miyagi}}, \bibinfo {author} {\bibfnamefont {J.~C.}\
  \bibnamefont {Berengut}}, \bibinfo {author} {\bibfnamefont {J.}~\bibnamefont
  {Biero{\'n}}}, \bibinfo {author} {\bibfnamefont {K.}~\bibnamefont {Blaum}},
  \bibinfo {author} {\bibfnamefont {L.~S.}\ \bibnamefont {Dreissen}}, \bibinfo
  {author} {\bibfnamefont {S.}~\bibnamefont {Eliseev}}, \bibinfo {author}
  {\bibfnamefont {P.}~\bibnamefont {Filianin}}, \bibinfo {author}
  {\bibfnamefont {M.}~\bibnamefont {Filzinger}}, \bibinfo {author}
  {\bibfnamefont {E.}~\bibnamefont {Fuchs}}, \bibinfo {author} {\bibfnamefont
  {H.~A.}\ \bibnamefont {F{\"u}rst}}, \bibinfo {author} {\bibfnamefont
  {G.}~\bibnamefont {Gaigalas}}, \bibinfo {author} {\bibfnamefont
  {Z.}~\bibnamefont {Harman}}, \bibinfo {author} {\bibfnamefont
  {J.}~\bibnamefont {Herkenhoff}}, \bibinfo {author} {\bibfnamefont
  {N.}~\bibnamefont {Huntemann}}, \bibinfo {author} {\bibfnamefont {C.~H.}\
  \bibnamefont {Keitel}}, \bibinfo {author} {\bibfnamefont {K.}~\bibnamefont
  {Kromer}}, \bibinfo {author} {\bibfnamefont {D.}~\bibnamefont {Lange}},
  \bibinfo {author} {\bibfnamefont {A.}~\bibnamefont {Rischka}}, \bibinfo
  {author} {\bibfnamefont {C.}~\bibnamefont {Schweiger}}, \bibinfo {author}
  {\bibfnamefont {A.}~\bibnamefont {Schwenk}}, \bibinfo {author} {\bibfnamefont
  {N.}~\bibnamefont {Shimizu}}, \ and\ \bibinfo {author} {\bibfnamefont
  {T.~E.}\ \bibnamefont {Mehlst{\"a}ubler}},\ }\href {\doibase
  10.1103/PhysRevLett.134.063002} {\bibfield  {journal} {\bibinfo  {journal}
  {\prl}\ }\textbf {\bibinfo {volume} {134}},\ \bibinfo {pages} {063002}
  (\bibinfo {year} {2025})}\BibitemShut {NoStop}%
\bibitem [{\citenamefont {Figueroa}\ \emph {et~al.}(2022)\citenamefont
  {Figueroa}, \citenamefont {Berengut}, \citenamefont {Dzuba}, \citenamefont
  {Flambaum}, \citenamefont {Budker},\ and\ \citenamefont
  {Antypas}}]{figueroa22prl}%
  \BibitemOpen
  \bibfield  {author} {\bibinfo {author} {\bibfnamefont {N.~L.}\ \bibnamefont
  {Figueroa}}, \bibinfo {author} {\bibfnamefont {J.~C.}\ \bibnamefont
  {Berengut}}, \bibinfo {author} {\bibfnamefont {V.~A.}\ \bibnamefont {Dzuba}},
  \bibinfo {author} {\bibfnamefont {V.~V.}\ \bibnamefont {Flambaum}}, \bibinfo
  {author} {\bibfnamefont {D.}~\bibnamefont {Budker}}, \ and\ \bibinfo {author}
  {\bibfnamefont {D.}~\bibnamefont {Antypas}},\ }\href@noop {} {\bibfield
  {journal} {\bibinfo  {journal} {\prl}\ }\textbf {\bibinfo {volume} {128}},\
  \bibinfo {pages} {073001} (\bibinfo {year} {2022})}\BibitemShut {NoStop}%
\bibitem [{\citenamefont {Ono}\ \emph {et~al.}(2022)\citenamefont {Ono},
  \citenamefont {Sato}, \citenamefont {Ishiyama}, \citenamefont {Higomoto},
  \citenamefont {Takano}, \citenamefont {Takasu}, \citenamefont {Yamamoto},
  \citenamefont {Tanaka},\ and\ \citenamefont {Takahashi}}]{ono22prx}%
  \BibitemOpen
  \bibfield  {author} {\bibinfo {author} {\bibfnamefont {K.}~\bibnamefont
  {Ono}}, \bibinfo {author} {\bibfnamefont {Y.}~\bibnamefont {Sato}}, \bibinfo
  {author} {\bibfnamefont {T.}~\bibnamefont {Ishiyama}}, \bibinfo {author}
  {\bibfnamefont {T.}~\bibnamefont {Higomoto}}, \bibinfo {author}
  {\bibfnamefont {T.}~\bibnamefont {Takano}}, \bibinfo {author} {\bibfnamefont
  {Y.}~\bibnamefont {Takasu}}, \bibinfo {author} {\bibfnamefont
  {Y.}~\bibnamefont {Yamamoto}}, \bibinfo {author} {\bibfnamefont
  {M.}~\bibnamefont {Tanaka}}, \ and\ \bibinfo {author} {\bibfnamefont
  {Y.}~\bibnamefont {Takahashi}},\ }\href@noop {} {\bibfield  {journal}
  {\bibinfo  {journal} {\prx}\ }\textbf {\bibinfo {volume} {12}},\ \bibinfo
  {pages} {021033} (\bibinfo {year} {2022})}\BibitemShut {NoStop}%
\bibitem [{\citenamefont {Wilzewski}\ \emph {et~al.}(2025)\citenamefont
  {Wilzewski}, \citenamefont {Spie\ss{}}, \citenamefont {Wehrheim},
  \citenamefont {Chen}, \citenamefont {King}, \citenamefont {Micke},
  \citenamefont {Filzinger}, \citenamefont {Steinel}, \citenamefont
  {Huntemann}, \citenamefont {Benkler}, \citenamefont {Schmidt}, \citenamefont
  {Huber}, \citenamefont {Flannery}, \citenamefont {Matt}, \citenamefont
  {Stadler}, \citenamefont {Oswald}, \citenamefont {Schmid}, \citenamefont
  {Kienzler}, \citenamefont {Home}, \citenamefont {Craik}, \citenamefont
  {Door}, \citenamefont {Eliseev}, \citenamefont {Filianin}, \citenamefont
  {Herkenhoff}, \citenamefont {Kromer}, \citenamefont {Blaum}, \citenamefont
  {Yerokhin}, \citenamefont {Valuev}, \citenamefont {Oreshkina}, \citenamefont
  {Lyu}, \citenamefont {Banerjee}, \citenamefont {Keitel}, \citenamefont
  {Harman}, \citenamefont {Berengut}, \citenamefont {Viatkina}, \citenamefont
  {Gilles}, \citenamefont {Surzhykov}, \citenamefont {Rosner}, \citenamefont
  {Crespo L\'opez-Urrutia}, \citenamefont {Richter}, \citenamefont {Mariotti},\
  and\ \citenamefont {Fuchs}}]{wilzewski25prl}%
  \BibitemOpen
  \bibfield  {author} {\bibinfo {author} {\bibfnamefont {A.}~\bibnamefont
  {Wilzewski}}, \bibinfo {author} {\bibfnamefont {L.~J.}\ \bibnamefont
  {Spie\ss{}}}, \bibinfo {author} {\bibfnamefont {M.}~\bibnamefont {Wehrheim}},
  \bibinfo {author} {\bibfnamefont {S.}~\bibnamefont {Chen}}, \bibinfo {author}
  {\bibfnamefont {S.~A.}\ \bibnamefont {King}}, \bibinfo {author}
  {\bibfnamefont {P.}~\bibnamefont {Micke}}, \bibinfo {author} {\bibfnamefont
  {M.}~\bibnamefont {Filzinger}}, \bibinfo {author} {\bibfnamefont {M.~R.}\
  \bibnamefont {Steinel}}, \bibinfo {author} {\bibfnamefont {N.}~\bibnamefont
  {Huntemann}}, \bibinfo {author} {\bibfnamefont {E.}~\bibnamefont {Benkler}},
  \bibinfo {author} {\bibfnamefont {P.~O.}\ \bibnamefont {Schmidt}}, \bibinfo
  {author} {\bibfnamefont {L.~I.}\ \bibnamefont {Huber}}, \bibinfo {author}
  {\bibfnamefont {J.}~\bibnamefont {Flannery}}, \bibinfo {author}
  {\bibfnamefont {R.}~\bibnamefont {Matt}}, \bibinfo {author} {\bibfnamefont
  {M.}~\bibnamefont {Stadler}}, \bibinfo {author} {\bibfnamefont
  {R.}~\bibnamefont {Oswald}}, \bibinfo {author} {\bibfnamefont
  {F.}~\bibnamefont {Schmid}}, \bibinfo {author} {\bibfnamefont
  {D.}~\bibnamefont {Kienzler}}, \bibinfo {author} {\bibfnamefont
  {J.}~\bibnamefont {Home}}, \bibinfo {author} {\bibfnamefont {D.~P. L.~A.}\
  \bibnamefont {Craik}}, \bibinfo {author} {\bibfnamefont {M.}~\bibnamefont
  {Door}}, \bibinfo {author} {\bibfnamefont {S.}~\bibnamefont {Eliseev}},
  \bibinfo {author} {\bibfnamefont {P.}~\bibnamefont {Filianin}}, \bibinfo
  {author} {\bibfnamefont {J.}~\bibnamefont {Herkenhoff}}, \bibinfo {author}
  {\bibfnamefont {K.}~\bibnamefont {Kromer}}, \bibinfo {author} {\bibfnamefont
  {K.}~\bibnamefont {Blaum}}, \bibinfo {author} {\bibfnamefont {V.~A.}\
  \bibnamefont {Yerokhin}}, \bibinfo {author} {\bibfnamefont {I.~A.}\
  \bibnamefont {Valuev}}, \bibinfo {author} {\bibfnamefont {N.~S.}\
  \bibnamefont {Oreshkina}}, \bibinfo {author} {\bibfnamefont {C.}~\bibnamefont
  {Lyu}}, \bibinfo {author} {\bibfnamefont {S.}~\bibnamefont {Banerjee}},
  \bibinfo {author} {\bibfnamefont {C.~H.}\ \bibnamefont {Keitel}}, \bibinfo
  {author} {\bibfnamefont {Z.}~\bibnamefont {Harman}}, \bibinfo {author}
  {\bibfnamefont {J.~C.}\ \bibnamefont {Berengut}}, \bibinfo {author}
  {\bibfnamefont {A.}~\bibnamefont {Viatkina}}, \bibinfo {author}
  {\bibfnamefont {J.}~\bibnamefont {Gilles}}, \bibinfo {author} {\bibfnamefont
  {A.}~\bibnamefont {Surzhykov}}, \bibinfo {author} {\bibfnamefont {M.~K.}\
  \bibnamefont {Rosner}}, \bibinfo {author} {\bibfnamefont {J.~R.}\
  \bibnamefont {Crespo L\'opez-Urrutia}}, \bibinfo {author} {\bibfnamefont
  {J.}~\bibnamefont {Richter}}, \bibinfo {author} {\bibfnamefont
  {A.}~\bibnamefont {Mariotti}}, \ and\ \bibinfo {author} {\bibfnamefont
  {E.}~\bibnamefont {Fuchs}},\ }\href {\doibase 10.1103/PhysRevLett.134.233002}
  {\bibfield  {journal} {\bibinfo  {journal} {\prl}\ }\textbf {\bibinfo
  {volume} {134}},\ \bibinfo {pages} {233002} (\bibinfo {year}
  {2025})}\BibitemShut {NoStop}%
\bibitem [{\citenamefont {Berengut}\ \emph {et~al.}(2020)\citenamefont
  {Berengut}, \citenamefont {Delaunay}, \citenamefont {Geddes},\ and\
  \citenamefont {Soreq}}]{berengut20prr}%
  \BibitemOpen
  \bibfield  {author} {\bibinfo {author} {\bibfnamefont {J.~C.}\ \bibnamefont
  {Berengut}}, \bibinfo {author} {\bibfnamefont {C.}~\bibnamefont {Delaunay}},
  \bibinfo {author} {\bibfnamefont {A.}~\bibnamefont {Geddes}}, \ and\ \bibinfo
  {author} {\bibfnamefont {Y.}~\bibnamefont {Soreq}},\ }\href@noop {}
  {\bibfield  {journal} {\bibinfo  {journal} {\prr}\ }\textbf {\bibinfo
  {volume} {2}},\ \bibinfo {pages} {043444} (\bibinfo {year}
  {2020})}\BibitemShut {NoStop}%
\bibitem [{\citenamefont {Miyake}\ \emph {et~al.}(2019)\citenamefont {Miyake},
  \citenamefont {Pisenti}, \citenamefont {Elgee}, \citenamefont {Sitaram},\
  and\ \citenamefont {Campbell}}]{miyake19prr}%
  \BibitemOpen
  \bibfield  {author} {\bibinfo {author} {\bibfnamefont {H.}~\bibnamefont
  {Miyake}}, \bibinfo {author} {\bibfnamefont {N.~C.}\ \bibnamefont {Pisenti}},
  \bibinfo {author} {\bibfnamefont {P.~K.}\ \bibnamefont {Elgee}}, \bibinfo
  {author} {\bibfnamefont {A.}~\bibnamefont {Sitaram}}, \ and\ \bibinfo
  {author} {\bibfnamefont {G.~K.}\ \bibnamefont {Campbell}},\ }\href {\doibase
  10.1103/PhysRevResearch.1.033113} {\bibfield  {journal} {\bibinfo  {journal}
  {\prr}\ }\textbf {\bibinfo {volume} {1}},\ \bibinfo {pages} {033113}
  (\bibinfo {year} {2019})}\BibitemShut {NoStop}%
\bibitem [{\citenamefont {R\"oser}\ \emph {et~al.}(2024)\citenamefont
  {R\"oser}, \citenamefont {Padilla-Castillo}, \citenamefont {Ohayon},
  \citenamefont {Thomas}, \citenamefont {Truppe}, \citenamefont {Meijer},
  \citenamefont {Stellmer},\ and\ \citenamefont {Wright}}]{roser24pra}%
  \BibitemOpen
  \bibfield  {author} {\bibinfo {author} {\bibfnamefont {D.}~\bibnamefont
  {R\"oser}}, \bibinfo {author} {\bibfnamefont {J.~E.}\ \bibnamefont
  {Padilla-Castillo}}, \bibinfo {author} {\bibfnamefont {B.}~\bibnamefont
  {Ohayon}}, \bibinfo {author} {\bibfnamefont {R.}~\bibnamefont {Thomas}},
  \bibinfo {author} {\bibfnamefont {S.}~\bibnamefont {Truppe}}, \bibinfo
  {author} {\bibfnamefont {G.}~\bibnamefont {Meijer}}, \bibinfo {author}
  {\bibfnamefont {S.}~\bibnamefont {Stellmer}}, \ and\ \bibinfo {author}
  {\bibfnamefont {S.~C.}\ \bibnamefont {Wright}},\ }\href@noop {} {\bibfield
  {journal} {\bibinfo  {journal} {\pra}\ }\textbf {\bibinfo {volume} {109}},\
  \bibinfo {pages} {012806} (\bibinfo {year} {2024})}\BibitemShut {NoStop}%
\bibitem [{\citenamefont {Schelfhout}\ and\ \citenamefont
  {McFerran}(2021)}]{schelfhout21pra}%
  \BibitemOpen
  \bibfield  {author} {\bibinfo {author} {\bibfnamefont {J.~S.}\ \bibnamefont
  {Schelfhout}}\ and\ \bibinfo {author} {\bibfnamefont {J.~J.}\ \bibnamefont
  {McFerran}},\ }\href@noop {} {\bibfield  {journal} {\bibinfo  {journal}
  {\pra}\ }\textbf {\bibinfo {volume} {104}},\ \bibinfo {pages} {022806}
  (\bibinfo {year} {2021})}\BibitemShut {NoStop}%
\bibitem [{\citenamefont {Zhang}\ \emph {et~al.}(2024)\citenamefont {Zhang},
  \citenamefont {Tiwari}, \citenamefont {Ganesh}, \citenamefont {Singh},\ and\
  \citenamefont {Flambaum}}]{Zhang_PhysRevResearch.6.043106}%
  \BibitemOpen
  \bibfield  {author} {\bibinfo {author} {\bibfnamefont {S.}~\bibnamefont
  {Zhang}}, \bibinfo {author} {\bibfnamefont {B.~S.}\ \bibnamefont {Tiwari}},
  \bibinfo {author} {\bibfnamefont {S.}~\bibnamefont {Ganesh}}, \bibinfo
  {author} {\bibfnamefont {Y.}~\bibnamefont {Singh}}, \ and\ \bibinfo {author}
  {\bibfnamefont {V.~V.}\ \bibnamefont {Flambaum}},\ }\href {\doibase
  10.1103/PhysRevResearch.6.043106} {\bibfield  {journal} {\bibinfo  {journal}
  {Phys. Rev. Res.}\ }\textbf {\bibinfo {volume} {6}},\ \bibinfo {pages}
  {043106} (\bibinfo {year} {2024})}\BibitemShut {NoStop}%
\bibitem [{\citenamefont {Beloy}\ and\ \citenamefont
  {Derevianko}(2008)}]{beloy08pra1}%
  \BibitemOpen
  \bibfield  {author} {\bibinfo {author} {\bibfnamefont {K.}~\bibnamefont
  {Beloy}}\ and\ \bibinfo {author} {\bibfnamefont {A.}~\bibnamefont
  {Derevianko}},\ }\href@noop {} {\bibfield  {journal} {\bibinfo  {journal}
  {\pra}\ }\textbf {\bibinfo {volume} {78}},\ \bibinfo {pages} {032519}
  (\bibinfo {year} {2008})}\BibitemShut {NoStop}%
\bibitem [{\citenamefont {Beloy}\ \emph {et~al.}(2008)\citenamefont {Beloy},
  \citenamefont {Derevianko},\ and\ \citenamefont {Johnson}}]{beloy08pra}%
  \BibitemOpen
  \bibfield  {author} {\bibinfo {author} {\bibfnamefont {K.}~\bibnamefont
  {Beloy}}, \bibinfo {author} {\bibfnamefont {A.}~\bibnamefont {Derevianko}}, \
  and\ \bibinfo {author} {\bibfnamefont {W.~R.}\ \bibnamefont {Johnson}},\
  }\href@noop {} {\bibfield  {journal} {\bibinfo  {journal} {\pra}\ }\textbf
  {\bibinfo {volume} {77}},\ \bibinfo {pages} {012512} (\bibinfo {year}
  {2008})}\BibitemShut {NoStop}%
\bibitem [{Note1()}]{Note1}%
  \BibitemOpen
  \bibinfo {note} {We use the notation $\protect \mathcal {M}$ instead of the
  usual standard $\mu $ for the nuclear magnetic moment to avoid confusion with
  the reduced mass introduced in Eq.~\protect \eqref {eq:kp}.}\BibitemShut
  {Stop}%
\bibitem [{\citenamefont {Kahl}\ and\ \citenamefont
  {Berengut}(2019)}]{kahl19cpc}%
  \BibitemOpen
  \bibfield  {author} {\bibinfo {author} {\bibfnamefont {E.~V.}\ \bibnamefont
  {Kahl}}\ and\ \bibinfo {author} {\bibfnamefont {J.~C.}\ \bibnamefont
  {Berengut}},\ }\href@noop {} {\bibfield  {journal} {\bibinfo  {journal}
  {\cpc}\ }\textbf {\bibinfo {volume} {238}},\ \bibinfo {pages} {232} (\bibinfo
  {year} {2019})}\BibitemShut {NoStop}%
\bibitem [{\citenamefont {Porsev}\ \emph {et~al.}(1999)\citenamefont {Porsev},
  \citenamefont {Rakhlina},\ and\ \citenamefont {Kozlov}}]{porsev99pra}%
  \BibitemOpen
  \bibfield  {author} {\bibinfo {author} {\bibfnamefont {S.~G.}\ \bibnamefont
  {Porsev}}, \bibinfo {author} {\bibfnamefont {Y.~G.}\ \bibnamefont
  {Rakhlina}}, \ and\ \bibinfo {author} {\bibfnamefont {M.~G.}\ \bibnamefont
  {Kozlov}},\ }\href@noop {} {\bibfield  {journal} {\bibinfo  {journal} {\pra}\
  }\textbf {\bibinfo {volume} {60}},\ \bibinfo {pages} {2781} (\bibinfo {year}
  {1999})}\BibitemShut {NoStop}%
\bibitem [{\citenamefont {Dzuba}\ \emph {et~al.}(1987)\citenamefont {Dzuba},
  \citenamefont {Flambaum}, \citenamefont {Silvestrov},\ and\ \citenamefont
  {Sushkov}}]{dzuba87jpb}%
  \BibitemOpen
  \bibfield  {author} {\bibinfo {author} {\bibfnamefont {V.~A.}\ \bibnamefont
  {Dzuba}}, \bibinfo {author} {\bibfnamefont {V.~V.}\ \bibnamefont {Flambaum}},
  \bibinfo {author} {\bibfnamefont {P.~G.}\ \bibnamefont {Silvestrov}}, \ and\
  \bibinfo {author} {\bibfnamefont {O.}~\bibnamefont {Sushkov}},\ }\href@noop
  {} {\bibfield  {journal} {\bibinfo  {journal} {\jpb}\ }\textbf {\bibinfo
  {volume} {20}},\ \bibinfo {pages} {1399} (\bibinfo {year}
  {1987})}\BibitemShut {NoStop}%
\bibitem [{\citenamefont {Godun}\ \emph {et~al.}(2014)\citenamefont {Godun},
  \citenamefont {Nisbet-Jones}, \citenamefont {Jones}, \citenamefont {King},
  \citenamefont {Johnson}, \citenamefont {Margolis}, \citenamefont {Szymaniec},
  \citenamefont {Lea}, \citenamefont {Bongs},\ and\ \citenamefont
  {Gill}}]{godun14prl}%
  \BibitemOpen
  \bibfield  {author} {\bibinfo {author} {\bibfnamefont {R.~M.}\ \bibnamefont
  {Godun}}, \bibinfo {author} {\bibfnamefont {P.~B.~R.}\ \bibnamefont
  {Nisbet-Jones}}, \bibinfo {author} {\bibfnamefont {J.~M.}\ \bibnamefont
  {Jones}}, \bibinfo {author} {\bibfnamefont {S.~A.}\ \bibnamefont {King}},
  \bibinfo {author} {\bibfnamefont {L.~A.~M.}\ \bibnamefont {Johnson}},
  \bibinfo {author} {\bibfnamefont {H.~S.}\ \bibnamefont {Margolis}}, \bibinfo
  {author} {\bibfnamefont {K.}~\bibnamefont {Szymaniec}}, \bibinfo {author}
  {\bibfnamefont {S.~N.}\ \bibnamefont {Lea}}, \bibinfo {author} {\bibfnamefont
  {K.}~\bibnamefont {Bongs}}, \ and\ \bibinfo {author} {\bibfnamefont
  {P.}~\bibnamefont {Gill}},\ }\href@noop {} {\bibfield  {journal} {\bibinfo
  {journal} {\prl}\ }\textbf {\bibinfo {volume} {113}},\ \bibinfo {pages}
  {210801} (\bibinfo {year} {2014})}\BibitemShut {NoStop}%
\bibitem [{\citenamefont {Huntemann}\ \emph {et~al.}(2016)\citenamefont
  {Huntemann}, \citenamefont {Sanner}, \citenamefont {Lipphardt}, \citenamefont
  {\protect{Chr.} Tamm},\ and\ \citenamefont {Peik}}]{huntemann16prl}%
  \BibitemOpen
  \bibfield  {author} {\bibinfo {author} {\bibfnamefont {N.}~\bibnamefont
  {Huntemann}}, \bibinfo {author} {\bibfnamefont {C.}~\bibnamefont {Sanner}},
  \bibinfo {author} {\bibfnamefont {B.}~\bibnamefont {Lipphardt}}, \bibinfo
  {author} {\bibnamefont {\protect{Chr.} Tamm}}, \ and\ \bibinfo {author}
  {\bibfnamefont {E.}~\bibnamefont {Peik}},\ }\href@noop {} {\bibfield
  {journal} {\bibinfo  {journal} {\prl}\ }\textbf {\bibinfo {volume} {116}},\
  \bibinfo {pages} {063001} (\bibinfo {year} {2016})}\BibitemShut {NoStop}%
\bibitem [{\citenamefont {Yang}\ \emph {et~al.}(2023)\citenamefont {Yang},
  \citenamefont {Wang}, \citenamefont {Wilkins},\ and\ \citenamefont
  {Ruiz}}]{yang23ppnp}%
  \BibitemOpen
  \bibfield  {author} {\bibinfo {author} {\bibfnamefont {X.}~\bibnamefont
  {Yang}}, \bibinfo {author} {\bibfnamefont {S.}~\bibnamefont {Wang}}, \bibinfo
  {author} {\bibfnamefont {S.}~\bibnamefont {Wilkins}}, \ and\ \bibinfo
  {author} {\bibfnamefont {R.~G.}\ \bibnamefont {Ruiz}},\ }\href@noop {}
  {\bibfield  {journal} {\bibinfo  {journal} {Prog. Part. Nucl. Phys.}\
  }\textbf {\bibinfo {volume} {129}},\ \bibinfo {pages} {104005} (\bibinfo
  {year} {2023})}\BibitemShut {NoStop}%
\bibitem [{\citenamefont {Ohayon}\ \emph {et~al.}(2022)\citenamefont {Ohayon},
  \citenamefont {Hofs\"ass}, \citenamefont {Padilla-Castillo}, \citenamefont
  {Wright}, \citenamefont {Meijer}, \citenamefont {Truppe}, \citenamefont
  {Gibble},\ and\ \citenamefont {Sahoo}}]{ohayon22njp}%
  \BibitemOpen
  \bibfield  {author} {\bibinfo {author} {\bibfnamefont {B.}~\bibnamefont
  {Ohayon}}, \bibinfo {author} {\bibfnamefont {S.}~\bibnamefont {Hofs\"ass}},
  \bibinfo {author} {\bibfnamefont {J.~E.}\ \bibnamefont {Padilla-Castillo}},
  \bibinfo {author} {\bibfnamefont {S.~C.}\ \bibnamefont {Wright}}, \bibinfo
  {author} {\bibfnamefont {G.}~\bibnamefont {Meijer}}, \bibinfo {author}
  {\bibfnamefont {S.}~\bibnamefont {Truppe}}, \bibinfo {author} {\bibfnamefont
  {K.}~\bibnamefont {Gibble}}, \ and\ \bibinfo {author} {\bibfnamefont {B.~K.}\
  \bibnamefont {Sahoo}},\ }\href@noop {} {\bibfield  {journal} {\bibinfo
  {journal} {\njp}\ }\textbf {\bibinfo {volume} {24}},\ \bibinfo {pages}
  {123040} (\bibinfo {year} {2022})}\BibitemShut {NoStop}%
\bibitem [{\citenamefont {{International Atomic Energy Agency
  (IAEA)}}(2025)}]{IAEA_Nuclear_Moments}%
  \BibitemOpen
  \bibfield  {author} {\bibinfo {author} {\bibnamefont {{International Atomic
  Energy Agency (IAEA)}}},\ }\href
  {https://www-nds.iaea.org/relnsd/vcharthtml/VChartHTML.html} {\enquote
  {\bibinfo {title} {{Nuclear Moments Database}},}\ } (\bibinfo {year}
  {2025}),\ \bibinfo {note} {[Accessed: March 10, 2025]}\BibitemShut {NoStop}%
\end{thebibliography}%
